\documentclass[journal]{IEEEtran}
\usepackage{amsmath}

\usepackage{amsthm}
\usepackage{amssymb}
\usepackage{graphicx}
\usepackage{subcaption}
\usepackage{url}
\usepackage{epstopdf}
\usepackage{placeins}
\usepackage{epsfig}
\usepackage{amsfonts}
\usepackage{balance}
\usepackage{algorithm} 

\usepackage{algorithmic}
\usepackage{epsfig}
\usepackage{subcaption}
\theoremstyle{remark}
\newtheorem{remark}{\textbf{Remark}}

\newtheorem{claim}{\textbf{Claim}}

\newtheorem{definition}{Definition}

\newtheorem{example}{\textbf{Example}}

\newcommand{\mytilde}{\raise.17ex\hbox{$\scriptstyle\mathtt{‌​\sim}$}}
\begin{document}
	\title{Index Coded - NOMA in Vehicular Ad Hoc Networks}

	\author{Sreelakshmi P, \IEEEmembership{~Member,~IEEE}, Jesy Pachat, \IEEEmembership{~Member,~IEEE},  Anjana A. Mahesh,\IEEEmembership{~Student Member,~IEEE}, Deepthi P. P., \IEEEmembership{~Member,~IEEE}, and B. Sundar Rajan, \IEEEmembership{Fellow,~IEEE}
		\thanks{Sreelakshmi P, Jesy Pachat and Deepthi P. P. are with the Electronics and Communication Department, National Institute of Technology, Calicut, 673601, India (e-mail: sreelakshmi.pazhoor@gmail.com, jesypachat@gmail.com, deepthi@nitc.ac.in).}
		\thanks{A. A Mahesh and B. S. Rajan are with the Department of Electrical Communication Engineering, Indian Institute of Science, Bangalore, 560012, India (e-mail: anjanamahesh@iisc.ac.in, bsrajan@iisc.ac.in).}
	}%
	
	{}

	\maketitle
	
	\begin{abstract} 
		The demand for multimedia services is growing day by day in vehicular ad-hoc networks (VANETs), resulting in high spectral usage and network congestion. Non-orthogonal multiple access (NOMA) is a promising wireless communication technique to solve the problems related to spectral efficiency effectively. The index coding (IC) is a powerful method to improve spectral utilization, where a sender aims to satisfy the needs of multiple receivers with a minimum number of transmissions. By combining these two approaches, in this work, we propose a novel technique called index coded NOMA (IC-NOMA), where we apply NOMA techniques on index coded data to reduce the number of transmissions further. This work shows that the IC-NOMA system demands a specific design for index codes to reap the advantages of NOMA. We have done the feasibility analysis of the proposed method in a general scenario and proposed an index code design to integrate IC over NOMA for the best efficiency. Through detailed analytical studies it is validated that the proposed transmission system provides improved spectral efficiency and power saving compared to conventional IC systems.
	\end{abstract}
	
	
	\IEEEpeerreviewmaketitle
	
	\section{INTRODUCTION}
	
	Vehicular Adhoc Networks (VANETs) have become a developing platform in modern intelligent transport systems. The prime goal of VANETs is to provide road safety and infotainment services to the users. The VANET scenario where a set of users demand some popular multimedia content is called the popular content dissemination scenario \cite{PCD}. The popular content may include navigational information such as road safety, real-time traffic data, road accident information etc. When multiple vehicles request common data, they can collaboratively download the data from a roadside unit (RSU)  with reduced download completion time \cite{cdd}. But in a general scenario, users will be interested in some common information for navigation as well as multimedia data of their choice for entertainment. Hence this work assumes that vehicles have some common demands together with some distinct individual demands.  All the vehicles are assumed to be equipped with long-term evolution-vehicle (LTE-V) based communication \cite{V2X11} devices. LTE-V is a promising integrated solution for supporting V2X communications with low latency and high reliability. The vehicles can download data packets as they pass through the range of RSU. The vehicles may not be able to download the complete set of packets demanded by them when they pass the RSU, as this depends on the speed of the vehicle and the range of RSU. In \cite{cdd2} authors have discussed the scenario where the vehicles can share their acquired data and satisfy their demands through cooperative communication. In this work, we assume that there is no cooperative communication among the vehicles. Vehicles are unwilling to collaborate due to privacy and security issues or the lack of incentive.

	The exchange of data needs to be less time-consuming, and any mistake can result in casualties.
	The concepts of index coding (IC) techniques can be exploited for overcoming bandwidth limitations and for the fast distribution of data \cite{VIC1}. Index coding is a source coding variant first proposed by Birk and Kol. The participating users have some prior details, called side information
	\cite{ICSI}. Based on each user's side information, the source would send coded blocks of data that would jointly allow users to extract the requested information in a minimum number of transmissions.
	The side information may be a linear combination of message packets, called coded side information \cite{ICCSI}, \cite{ICCSI2}.
	The primary goal of an IC problem is to minimize the number of transmissions required and improve bandwidth efficiency. The index coding technique can be introduced to VANETs, which lets RSUs to serve the users with side information by reduced number of transmissions.
	The number of transmissions reduces, and thus the network load, which improves the overall transmission efficiency \cite{VIC1}. 
	
	V2X communicating units process and share huge amounts of data which can result in increased latency and irregular connectivity. 5G networking technologies built into vehicular networks satisfy the increasing communication needs. One of the most critical factors in improving network specifications is the implementation of an effective multiple access technique. Non-orthogonal multiple access (NOMA) techniques \cite{ON} are found to have increased spectral efficiency and decreased latency compared with orthogonal multiple access (OMA) techniques. By accommodating far more users than OMA,
	NOMA meets the demand for massive connectivity. NOMA supports different users using the same resources. NOMA introduces controllable interference to superpose different user signals at the expense of a tolerable increase in the complexity of the receivers. The channel gains of users vary depending on their distance from the transmitter, among other factors. Users with a substantial difference in channel gain need a single NOMA transmission with superposition coding (SC) \cite{N1}, \cite{N3}. The far users are assigned higher powers due to their weak channel conditions than near users with good channel conditions. The signals of far users are superposed linearly with signals of near users for transmission. A decoding process called successive interference cancellation (SIC) takes place at the receiver based on the distance from the sender \cite{N4}. The far-user must decode the data from the superposed signal by treating the interference from any other device signal as noise. The others must decode and deduct the highest power data from the received signal to decode the desired signal. Then all users are served concurrently by SC at transmitter and SIC at the receiver using available bandwidth. NOMA-based studies are evolving day in and day out. Allocating one frequency channel to multiple users at the same time makes NOMA a promising multiple access system with improved spectral efficiency.
	Cache-assisted NOMA (CA-NOMA) exploits cached and previously decoded data for better cancellation of interference called cache-enabled interference cancellation \cite{CN3}. In \cite{CNIC} authors showed that in a system with two users, when each user cached the data demanded by other user, then switching from SC to IC boost the power efficiency.
	
	\subsection{Contributions}
	In this work, we propose a transmission strategy in VANETs called index coded-NOMA by integrating the concept of index coding with that of NOMA. The side information of each vehicle can be exploited effectively to design index codes with a minimum number of transmissions from the sender. 
	Applying NOMA principles over index coding can further reduce the number of transmissions to satisfy the demands of users. 
	Here we consider a particular VANET scenario, where the vehicles are moving in a multilane track with almost equal speed.  Under this framework, we combine NOMA principles and IC techniques to develop a transmission strategy with superior spectral efficiency.

	IC-NOMA superposes index coded packets at the transmitter to serve more users simultaneously per transmission, thus minimizing the number of transmissions. But in order to get a spectral advantage through IC-NOMA, it is required to develop a  suitable index code design. Therefore, this work proposes an algorithmic solution to design index codes that fit with NOMA concepts. 
	The algorithm proposed requires a lesser number of transmissions from the server compared to conventional IC. We compare the performance of the proposed transmission algorithm with conventional IC analytically. It is seen that the proposed algorithm offers improved bandwidth efficiency along with power efficiency. 
	
	To the best of our knowledge, this is the first work that considers IC-NOMA. The main contributions of this paper are summarized below:
	\begin{itemize}
		\item[$\bullet$] We propose IC-NOMA as a transmission strategy with superior spectral efficiency in VANETs. 
		\item[$\bullet$] It is shown that IC-NOMA needs a unique index code design to get bandwidth improvement. This work develops an algorithm to design index codes for IC-NOMA.
		\item[$\bullet$] Through detailed analytical studies, it is showed that the proposed transmission system provides improved spectral efficiency and power saving compared to conventional IC systems.
	\end{itemize}

	\begin{figure}[]
		\centering
		\includegraphics[width=9cm, height=5cm]{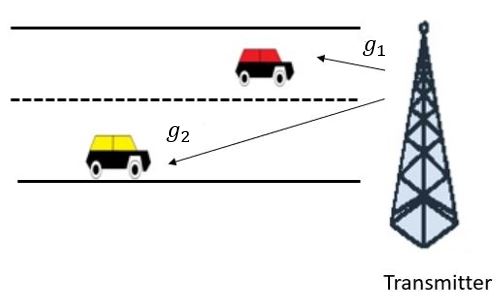}
		\caption{NOMA scenario in VANET with $M = 2$.}
		\label{fig:NOMA_M=2}
	\end{figure}
	\section{Preliminaries and Background}
	\label{prelims}
	In this section, we discuss some basics of NOMA and IC. These promising wireless techniques have become a research hotspot in vehicular communications.
	\subsection{Non-orthogonal multiple access (NOMA)}
	
	NOMA can be widely categorized into two, namely power domain NOMA and code domain NOMA. In power domain NOMA, different user signals are superposed on the power domain depending on their channel conditions.  
	The users served simultaneously by a single superposed transmission form a cluster \cite{N2}. 

	Assuming $M$ users per cluster and each user denoted as $V_{i}$ with $i\in [M]$, let $x_i$ be the data requested by user $V_i$. Without loss of generality, let user $V_{M}$ be the farthest user and $V_{1}$ be the nearest user, and the channel gains be $g_1 >g_2 >g_3 .....>g_M$. The signal $s_{i}$ be the encoded form of $x_{i}$, $s_{i}=enc(x_{i})$. Let the additive white gaussian noise with zero mean and variance $\sigma_{i}^{2}$ at $V_{i}$ be represented as $ n_{i}$. 
	
	Considering the channel gain of each user, the signals requested by them are linearly superposed in power domain at the transmistter as 
	\begin{equation}
	S=\sum_{i=1}^{M}\sqrt{\alpha_{i}P}s_{i},
	\end{equation}
	where $P$ is the transmission power, $\alpha_{i} P$ is the amount of power allocated to the signal corresponding to user $V_{i}$ with $\alpha_1 <\alpha_2 <\alpha_3 .....<\alpha_M$ and $\sum_{i=1}^{M}\alpha_{i} = 1$. 
	Then the received signal at $V_{i}$ is
	\begin{equation}
	Z_{i} =\sqrt{ g_{i}}S + n_{i}.
	\end{equation}    
	For user demultiplexing, SIC is carried out at near users \cite{N5}, based on the power with which the base station transmits the signals. The far user decodes the desired information, considering interference due to others as noise. All other users perform SIC, where the highest power data is decoded and progressively cancelled to decode the desired information.
	
	Fig.\ref{fig:NOMA_M=2} shows a conventional NOMA scenario with $M=2$.
	In this case, the transmitted signal is  $S = \sqrt{\alpha_{1}P} s_{1} + \sqrt{\alpha_{2}P} s_{2}$ with  $\alpha_{2}=1-\alpha_{1}$. Let $\alpha <0.5$ be the power allocation factor and $\alpha_{1} =\alpha$, $\alpha_{2}=1-\alpha$. The transmitted signal is given by $S = \sqrt{\alpha P} s_{1} + \sqrt{(1-\alpha)P} s_{2}$. Users $V_{1}$ and $V_{2}$ receives $Z_{1}=\sqrt{g_{1}}S + n_{1}$ and $Z_{2}=\sqrt{g_{2}}S + n_{2}$ respectively. Far user $V_{2}$ decodes $x_{2}$ by considering interference due to $x_{1}$ as noise, $V_{1}$ performs SIC of $x_{2}$ to decode $x_{1}$. After SIC the signal at $V_{1}$ can be represented as $Z'_{1}=Z_{1}-\sqrt{g_{1}}\sqrt{(1-\alpha)P} s'_{2}$ with  $s'_{2}$ as the signal intended for $V_{2}$ decoded at $V_{1}$.

	\subsection{Index coding}
	\label{P-C}
	
	Consider a single sender with $n$ messages $\mathcal{X}=\{x_{1}, x_{2},...x_{n}\}$, $x_i\in \mathbb{F}_q$, where $ \mathbb{F}_q$ denotes the finite field with $q$ elements for a prime power $q$. Let there be $N$ receivers each with some prior knowledge of these messages over a broadcast channel \cite{ICTP}. A receiver denoted as $V_{i}$, $i\in[N]$ needs a subset of the messages denoted as $\mathcal{W}_{i}$, with $\mathcal{W}_{i} \subseteq \mathcal{X}$, known as the want set. Also, each receiver is holding a subset of these messages termed as known set or side information and denoted as $\mathcal{K}_{i}$, with  $\mathcal{K}_{i} \subseteq \mathcal{X}$. Let the known set and want set of index coding problem is denoted as $\mathcal{K}= \{\mathcal{K}_{i}:  \quad i\in [N]\}$ and $\mathcal{W}= \{\mathcal{W}_{i}:  \quad i\in [N]\}$. The index coding problem with $n$ messages, $N$ receivers, known set $\mathcal{K}$ and want set $\mathcal{W}$ be denoted as $\mathcal{I}(n, N,\mathcal{K}, \mathcal{W})$.
	\begin{definition} For instance  $\mathcal{I}(n, N,\mathcal{K}, \mathcal{W})$ of an index coding problem with input vector $ \textbf{x}\in\mathbb{F}_{q}^{n}$, then the corresponding index code \cite{ICSI}, $\textbf{y}\in\mathbb{F}_{q}^{l}$ of  length  $l$,  consists of 
		\begin{enumerate}
			\item An encoding function $\mathcal{F}: \mathbb{F}_{q}^{n}  \to \mathbb{F}_{q}^{l}$.
			\item And corresponding decoding functions for each receiver $\mathcal{G}_{i}: \mathbb{F}_{q}^{l} \times  \mathbb{F}_{q}^{|\mathcal{K}_{i}|} \to \mathbb{F}_{q}^{|\mathcal{W}_{i}|} $ for $i \in [N]$.
			
		\end{enumerate}
	\end{definition}
	
%
If $l$ is minimum then the index code designed is optimal. For a linear index code $\textbf{y}$, the encoding function is a linear transformation described as, $\textbf{y}=\textbf{L}\textbf{x},$ where $\textbf{L} \in \mathbb{F}_{q}^{l\times n}$ is known as the encoding matrix and is defined as $\textbf{L} =\mathcal{F}(n, N,\mathcal{K}, \mathcal{W})$.

	In \cite{ICCSI}, \cite{ICCSI2} authors give a generalization of index coding problem, where the elements of known set are linear combinations of the messages. Such index coding problems are known as index coding with coded side information (ICCSI). The coded side information at receiver $V_{i}$ is represented using the vector, $\textbf{a}_{i} \in\mathbb{F}_{q}^{d_{i}}$ such that $\textbf{a}_{i} = \textbf{B}_{i}\textbf{x}^T$, where $\textbf{B}_{i} \in\mathbb{F}_{q}^{d_{i}\times n}$ is known as the side-information generator matrix.  Under this set up let $\textbf{L}_c$ be the encoding matrix used at the sender such that the demands of all users are satisfied in minimum number of transmissions. So the index coding vector for ICCSI problem is given as $\textbf{y}_c=\textbf{L}_c\textbf{x}$. Here  $\textbf{L}_c =\mathcal{F}(n, N,\mathcal{K}_c, \mathcal{W})$, where $\mathcal{K}_c=\{\textbf{a}_{i}:i\in [N]\}$ is the coded side information.
	
	\section{System model and Motivating Example}
	\label{sysmod}
	
	In this section, we discuss the system model for the proposed IC-NOMA scheme and a method to design index codes for IC-NOMA to achieve improved bandwidth efficiency.
	\subsection{System model}
	 This work considers a downlink VANET scenario in which vehicles travel on a one-way multi-lane road. The vehicles travel at speed such that the relative velocity between them is negligible.
		The system model of this work considers LTE-V or Cellular-V2X (C-V2X) \cite{cv2x} standard. As per the technical specifications published by 3GPP in 2016, this standard enable both direct and network-based communications. Thus, the vehicles can exchange information with infrastructures such as roadside units (RSU) or base stations (BS). The RSU and vehicles in its range can communicate directly over 5.9GHz unlicensed frequency, whilst the vehicles can interact with the base station using the C-V2X licensed spectrum.

	
		\begin{figure}[]
		\centering
		\includegraphics[width=9.5cm, height=6cm]{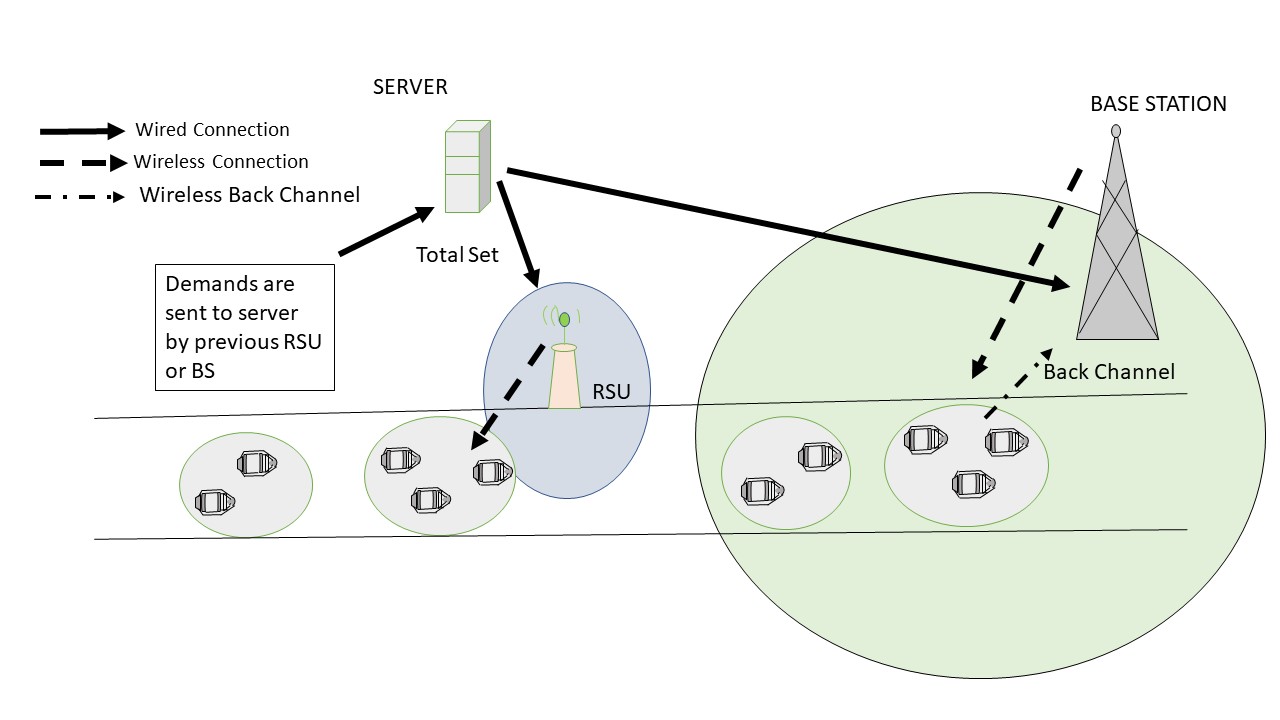}
		\caption{VANET scenario with $N = 5, N_f= 2$.}
		\label{fig:ICNOMA2}
	\end{figure}
	
	All the vehicles moving in one direction form a cluster. The vehicles in the cluster are assumed to be interested in some common information while each individual vehicle has some distinct demand for some entertainment data. Therefore the demand set of users can have some packets demanded in common. The vehicles report their demands to the server through RSU or BS. The server will send the entire set of demanded packets to the next RSU and BS in the direction of vehicle movement. The communication of the server with BS and RSU is achieved through wireline channels. RSU delivers the whole set of demanded packets to the cluster of vehicles when they enter its range. Each vehicle receives some of the packets as their side information. This work assumes that each vehicle has a cache where data packets delivered by RSU nodes are stored as side information to reduce the cost associated with network usage. As the vehicles are not entering the RSU range simultaneously, their side information could be different when they leave the range of RSU.  
	The vehicles will send the index set of their side information to BS through a wireless backchannel. The BS performs IC-NOMA to the users in its range.

	 Hence the scenario under consideration consists of different phases as follows:
	\begin{enumerate}
		\item Reporting Phase:  The vehicles can report their demands to server through  RSU or BS. The server on receiving user demands send the total demand set to the next RSU and base station (BS) in the forward direction of the vehicles through the respective wireline channels.
		\item R2V Phase: The RSU disseminate data packets to users in its transmission range, and each user receives subset of the packets delivered by RSU as their side information.  
		\item V2N Phase:  In this phase vehicles communicate only to the BS. The vehicles communicate the index set of their side information to the BS through a wireless back channel. The BS updates the demand set of vehicles by removing the messages already received by them in R2V phase. The base station performs IC-NOMA for users entering the communication range of BS. 
	\end{enumerate}

	\subsection{IC-NOMA in VANET}
	In this section, we introduce the concepts of the proposed transmission strategy with superior spectral efficiency called  IC-NOMA, which is an integration of NOMA principles and index coding techniques. Unlike traditional NOMA, where message packets are superposition coded, IC-NOMA superposes index coded packets in the power domain. 
	
Consider a cluster of $N$ vehicles with $\mathcal{D}_{i}$, $i \in [N]$ as the initial set of demands which the vehicle $V_i$ communicates to the server via RSU or BS.
 The server sends the whole set of demanded packets denoted as, $\mathcal{X}=\bigcup_{i\in [N]} \mathcal{D}_{i}$ to the RSU and BS in the direction of vehicle movement. During R2V phase, $V_i$ will receive their side information packets denoted as $\mathcal{K}_{i}$. For the IC problem under consideration, the demands of each user is modified at BS as $\mathcal{W}_{i}=\mathcal{D}_{i} \backslash \{\mathcal{D}_{i} \cap \mathcal{K}_{i}\}$ to remove the demanded messages that are received as side information during R2V phase.
	
		Let $g_i$, $i\in [N]$ denote the  channel gain between the BS and the receiving vehicle $V_i$. Assume there is a group of $N_f$ far users and a group of $N_n =N-N_f$ near users, with approximately same channel gains for users within a group and a significant channel gain difference between users in different groups. The channel gains of users are denoted as $g_1\approxeq g_2......\approxeq g_{N_n} > g_{N_n+1} ......\approxeq g_{N}$. Fig. \ref{fig:ICNOMA2} represents VANET scenario under consideration with $N=5$, $N_f=2$ and $N_n=3$.

	The minimum number of transmissions needed for the IC-NOMA scenario is represented by $l^{\scriptscriptstyle IC-NOMA}$ and that for conventional IC  is represented by $l^{\scriptscriptstyle IC}$. Later we prove that $l^{\scriptscriptstyle IC-NOMA} \le l^{\scriptscriptstyle IC}$.
	
	The $j^{th}$ IC-NOMA signal received by user $V_i$ can be represented as  
	\begin{align}
	\centering
	Z_i^{j}&= \sqrt{g_i}S_{j} + n_i, \quad i\in [N], \hspace{0.1cm} j \in [l^{\scriptscriptstyle IC-NOMA}], 
	\end{align}
where $S_{j}$ with $j \in [l^{\scriptscriptstyle IC-NOMA}]$ represents the $j^{th}$ transmitted signal, $g_i$ for $i\in[N]$ represents the channel gain between the user $V_i$ and the BS and $n_i$ represents the additive white Gaussian noise with zero mean and variance $\sigma_{i}^2$ at user $V_i$.

	\subsection{Motivating Example}
	
	In this section we show the proposed transmission stratergy  called IC-NOMA through an example. 
	\begin{example}
		\label{ex:ICN1}
		Let $N=3$, Table \ref{ICNOMAexp1} displays known set and corresponding want set of each vehicle. 
	The linear solution for the given IC problem requires atleast two transmissions. One such optimal linear index code is given by the encoding matrix $\textbf{L}$ as
 
		$$\textbf{L}=
		\begin{bmatrix}
		1 & 1 & 0 \\
	 	0 & 0 & 1
		\end{bmatrix}.
		$$ 
	 The corresponding index coded transmissions are $y_{1}=x_{1}+x_{2}$ and $y_{2}=x_{3}$. 
		
		In this example $V_{1}, V_{2}$ are near users and $V_{3}$ is the far user so that $N_f=1$ and $N_n=2$. Consider  the channel gains of users as $g_1\approxeq g_2> g_3$. The IC-NOMA transmission combines the concepts of NOMA with IC to transmit the signals $s_{1}=enc(y_{1})$ and $s_{2}=enc(y_{2})$. 
		
		Hence the IC-NOMA transmitted signal in this case is $S = \sqrt{\alpha P} s_{1} + \sqrt{(1-\alpha)P} s_{2}$ with $\alpha <0.5$.
		Here data requested by two near users $V_{1}, V_{2}$ are paired by index coding under the same power level. 
		The far user $V_3$ gets $x_3$ transmitted at highest power, near users $V_1$ and $V_2$ decode $y_{2}=x_3$ and perform SIC to get the index coded packet $y_1=x_{1}+x_{2}$. They exploit the side information to decode the desired data.
		Eventually, the designed scheme is equivalent to a two-user NOMA system with two power levels, though there are three users. Hence the number of users served simultaneously by a single transmission increases.
		
	\end{example}
	\begin{table}[h]
		\centering
		\caption{Example 1 of IC-NOMA problem}
		\begin{tabular}{|c|c|c|}
			\hline
			vehicle& {Known set $\mathcal{K}_i$}  & {Want set $\mathcal{W}_i$}\\ \hline
			$V_1 $      & $\{x_2\} $ &$\{ x_1\} $ 
			\\ \hline
			$    V_2 $   & $ \{x_1\}$   & $\{x_2\} $ \\ \hline
			$    V_3 $   & &$ \{x_3\} $  \\ \hline
			
		\end{tabular}
		\label{ICNOMAexp1}
	\end{table}
	The total number of required transmissions in IC-NOMA for Example \ref{ICNOMAexp1} is $ l^{\scriptscriptstyle IC-NOMA} =1$. 
	\subsection{Index Code Design for IC-NOMA}
	In a traditional index coding problem, optimal linear index code can be designed based on want set and known set of each user. But some IC solutions cannot be combined with NOMA principles efficiently even when the index code is optimal. Hence it is required to see how index code design needs be modified to reap additional power or bandwidth saving when combined with NOMA. In NOMA decoding, near user apply SIC for decoding and can retrieve the information intended for both the near and far user (ie., the data sent at both the power levels). So in IC-NOMA, index coded data sent to the far user can be used as an additional coded side information to meet the demands of the near user.
	 In contrast, the demands of the far users should be satisfied with the index coded packets transmitted at the higher power, because the index coded packets sent at the lower power levels are considered as interfering noise and are discarded. Therefore, the signals at the highest power levels should provide a complete IC solution to meet the demands of far users. Hence an index code design that considers the information redistribution through NOMA decoding need to be developed for achieving improved transmission efficiency.  
	
	Let $x = [x_1\hspace{0.1cm} x_2 ...x_n]^T$ be the set of messages to be communicated. 
	Considering only the want set and known set of far users denoted as $\mathcal{W}_{f} = \{\mathcal{W}_{i}:  \quad i\in \mathcal{I}_{f}\}$ and  $\mathcal{K}_{f}=\{ \mathcal{K}_{i}:  \quad i\in \mathcal{I}_{f}\} $ with $\mathcal{I}_{f}$ as the index set of far users, a linear IC problem is formulated. A linear index code specified by the encoding matrix $\textbf{L}_f =\mathcal{F}(n,|\mathcal{I}_{f}| ,\mathcal{K}_f, \mathcal{W}_f)$ is given by
	\begin{equation}
	\label{eq3}
	\textbf{y}_f=\textbf{L}_f\textbf{x},
	\end{equation} 
	
	\noindent where $\textbf{L}_f \in \mathbb{F}_{q}^{l_f\times n}$, the length of index code designed for far users is denoted as $l_f$. Let $\mathcal{Y}_f=\{y_{f_i}: \quad i\in [l_f]\}$ be the set of index coded packets designed for far users. 
	The near users decode the data of far users to perform SIC; hence an index code is designed for near users by exploiting the respective side information along with the index code of far users as coded side information.
	
	  Consider $\mathcal{I}_{n}$ as the index set of near users. The side information set at each near user can be modified as $\mathcal{K}_{i}'= \{\mathcal{K}_{i}\cup \{ \mathcal{Y}_f \}: \quad i\in \mathcal{I}_{n}\}$ and $\mathcal{K}_{n}' = \{\mathcal{K}_{i}':  \quad i\in \mathcal{I}_{n}\} $. The total want set of near users are represented as $\mathcal{W}_{n}'=\{ \mathcal{W}_{i}\backslash\mathcal{W}_{i}^c:  \quad i\in \mathcal{I}_{n}\} $, where $\mathcal{W}_{i}^c$ is the demand set of near user $V_i$ satisfied through coded side information of far users.  Hence the index coding problem for near users is an ICCSI problem with encoding matrix $\textbf{L}_n^c =\mathcal{F}(n,|\mathcal{I}_{n}| ,\mathcal{K}_n', \mathcal{W}_n')$. The index coded vector is obtained as,
	
	\begin{equation}
	\label{eq4}
	\textbf{y}_n^c=\textbf{L}_n^c\textbf{x},
	\end{equation} 
	
	\noindent where  $\textbf{L}_n^c \in \mathbb{F}_{q}^{l_n\times n}$,  and $l_n$ is the length of index code designed for near users.

	The side information of each user, along with the distance from the BS, significantly influence the design of IC-NOMA. This is illustrated through an example.
	\begin{example}
		\label{ex:ICN2}
		Consider the VANET scenario shown in Fig. \ref{fig:ICNOMA2} with $N=5$.
		Table \ref{ICNOMAexp2} displays the want set of each vehicle.\\
		
		\begin{table}[h]
		
			\centering
			\caption{Users with their demands(Before R2V Phase).}
			\begin{tabular}{|c|c|c|}
				\hline
				User	&  Demands ($\mathcal{D}_i$)  \\
				\hline
				$V_{1}$	& $\{x_1, x_4, x_5, x_6\}$ \\
				\hline
				$V_{2}$	&$\{ x_1, x_5, x_6\}$  \\
				\hline
				$V_{3}$	& $\{x_1, x_2, x_6\}$ \\
				\hline
				$V_{4}$	&$\{x_1, x_3, x_4, x_5, x_7\}$ \\
				\hline
				$V_{5}$	& $\{x_3, x_6\}$\\
				\hline
			\end{tabular}
			\label{ICNOMAexp2}
		\end{table}
		The server sends the total demand set $\mathcal{X}=\{x_1,x_2,x_3,x_4,x_5,x_6,x_7\}$ to RSU in the direction of vehicle movement. The vehicles on entering the RSU range receives data packets as side information as shown by 	Table \ref{ICNOMAexp3}. Since the demands of some of the users will be satisfied by side information during R2V phase, the want set is modified at BS as $\mathcal{W}_{i}=\mathcal{D}_{i} \backslash \{\mathcal{D}_{i} \cap \mathcal{K}_{i}\}$,   Table \ref{ICNOMAexp3} shows the modified want set.
		
		 Here it is assumed that spacing between vehicles $V_1$ and $V_2$ is same as that between $V_2$ and $V_3$ as well as $V_4$ and $V_5$ while spacing between $V_3$ and $V_4$ is longer. Since the overlap in side information between consecutive vehicles depends on spacing between them, the overlap  in side information between $V_3$ and $V_4$ is lesser. Also since $V_5$ enters last, it could download lesser number of packets compared to other vehicles.
		\begin{table}[h]
			\centering
				\caption{Example 2 of IC-NOMA problem}
			\begin{tabular}{|c|c|c|}
				\hline
				User	& Known set  & want set $\mathcal{W}_i$ (After R2V Phase) \\
				\hline
				$V_{1}$	& $\{x_1,x_2,x_3\}$ & $\{x_4, x_5, x_6\}$ \\
				\hline
				$V_{2}$	& $\{x_2,x_3,x_4\}$ &$\{x_1, x_5, x_6 \}$  \\
				\hline
				$V_{3}$	& $\{x_3,x_4,x_5\}$ &  $\{ x_1, x_2, x_6\}$ \\
				\hline
				$V_{4}$	&$\{x_5,x_6,x_7\}$  &$\{x_1, x_3, x_4\}$  \\
				\hline
				$V_{5}$	& $\{x_6,x_7\}$ &  $\{x_3\}$\\
				\hline
			\end{tabular}
		
			\label{ICNOMAexp3}
		\end{table}
		
		The encoding matrix representing one of the possible linear solutions for the given index coding problem is
		$$\textbf{L}=
		\begin{bmatrix}
		1 & 0 & 0 & 1 & 0 & 0 & 0 \\
		0 & 1 & 0 & 0 & 1 &0 & 0\\
		0 & 0 & 1 & 0 & 0 & 1 & 0  \\
		0 & 0 & 0 & 1 & 0 & 0 & 1	 \\
		\end{bmatrix}.
		$$

	
		The index coded transmissions designed are $y_{1}=x_{1}+x_{4}$, 
		$y_{2}=x_{2}+x_{5}$, $y_{3}=x_{3}+x_{6}$, 
		$y_4=x_4+x_7$. 	The index code length is given by $l^{\scriptscriptstyle IC}=4$.\\
		
			\begin{table}[]
			\centering
			\caption{Example 2 of IC-NOMA problem for far users}
			\begin{tabular}{|c|c|c|}
				\hline
				vehicle & {Known set}& {want set} \\ \hline
				$V_{4}$	&$\{x_5, x_6, x_7\} $ &$\{x_1, x_3, x_4\}$  \\
				\hline
				$V_{5}$	& $\{x_6,x_7\}$ &  $\{x_3\}$\\ \hline
			\end{tabular}
			
			\label{ex2f}
		\end{table}
		
		\begin{table}[]
			\centering
			\caption{Example 2 of IC-NOMA problem for near users with modified want set and known set.}
			\begin{tabular}{|c|c|c|}
				\hline
				vehicle & { Known set}& {want set} \\ \hline
				$V_{1}$	& $\{x_1,x_2,x_3, x_{1}+x_{7},  x_{3}+x_{6},  x_{4}+x_{7}\}$ & $\{x_5\}$ \\
				\hline
				$V_{2}$	& $\{x_2,x_3,x_4,  x_{1}+x_{7},  x_{3}+x_{6},  x_{4}+x_{7}\}$ &$\{x_5\}$  \\
				\hline
				$V_{3}$	& $\{x_3,x_4,x_5,  x_{1}+x_{7},  x_{3}+x_{6},  x_{4}+x_{7}\}$ &  $\{x_2\}$ \\
				\hline
			\end{tabular}
			
			\label{ex2n}
		\end{table}
		When combining the designed index codes with NOMA principles, the far users should be able to decode the desired data without getting the data of near users. We cannot use the above index code design in the NOMA scenario, as the far users need index coded packets of near users for decoding the desired data. Hence in an IC-NOMA scenario, the index code design need to be modified for combining with NOMA.
		
		In this example, as described earlier spacing between $V_3$ and $V_4$ is higher than others. Therefore $V_1$, $V_2$ and $V_3$ are grouped as near users while $V_4$ and $V_5$ are grouped as far users. Thus $N_f=2$ and $N_n=3$ in this case. The channel gains of users are  $ g_1\approxeq  g_2 \approxeq g_3 > g_4\approxeq g_5$.

		In IC-NOMA, to develop index codes that can bring in improved transmission efficiency when combined with NOMA scheme, we consider the index coding problem as two different index coding problems: One as a conventional IC problem to design index codes for far users; and the other as an ICCSI for near users.

		First the encoding matrix $\textbf{L}_f$ is designed to develop index codes for far users as per (\ref{eq3}) based on Table \ref{ex2f} for far users. Hence the encoding matrix $\textbf{L}_f$ corresponding to far users is designed as
		$$\textbf{L}_f=
		\begin{bmatrix}
		1 & 0 & 0 & 0 & 0 & 0 & 1  \\
		0 & 0 & 1 & 0 & 0 & 1 & 0 \\
		0 & 0 & 0 & 1 & 0 & 0 & 1 
		\end{bmatrix}.
		$$

		The index code length for far users is $l_f=3$ and is given by $y_{f_{1}}= x_{1}+x_{7}$, $y_{f_{2}}= x_{3}+x_{6}$ and $y_{f_{3}}= x_{4}+x_{7}$. 
		
		Here some of the near user demands are satisfied by the index code of far user. Hence the want set along with the known set of near users are modified as shown in Table \ref{ex2n}.
		
		As per (\ref{eq4}) index coded solutions can be developed for near users based on Table \ref{ex2n}. Using $y_{f_{1}}$, $ y_{f_{2}}$ and $ y_{f_{3}}$ as coded side information along with the respective side information, we can design an encoding matrix $\textbf{L}_n^c$ for near users.
		$$\textbf{L}_n^c=
		\begin{bmatrix}
		0 & 1 & 0 & 0 & 1 & 0 & 0
		\end{bmatrix}.
		$$ 
		
	\noindent	$y_{n_1}^c$ =	$x_2+x_5$ are the index code designed for near users $V_{1}, V_{2}$ and $V_{3}$ with length as $l_n=1$.

		The index codes designed are combined using NOMA principles in IC-NOMA.
		Hence the transmitted signals in IC-NOMA is $S_1=\sqrt{\alpha P} s_{n_1} + \sqrt{(1-\alpha)P} s_{f_1}$, $S_2=\sqrt{ P}s_{f_2}, S_3=\sqrt{ P}s_{f_3}$,\\
		where  $\alpha < 0.5$ and $s_{n_1}=enc(y_{n_1}^c)$, $s_{f_1}=enc(y_{f_1})$, $s_{f_2}=enc(y_{f_2})$ and $s_{f_3}=enc(y_{f_3})$. 
		Hence the number of required transmissions, in this case, is $ l^{\scriptscriptstyle IC-NOMA} =3$.
		
		The far users receiving $S_1$ decode $s_{f_1}$  considering interference due to $s_{n_1}$ as noise, and from $S_2$ and $S_3$ they decode $s_{f_2}$ and $s_{f_3}$.  
		The near users receiving $S_1$ get the  far user data $s_{f_1}$ perform SIC to get $s_{n_1}$, and  from $S_2$ and $S_3$ they decode $s_{f_2}$ and $s_{f_3}$.	
		Hence the far users get $\{s_{f_1}, s_{f_2}, s_{f_3}\}$, whereas the near users get $\{s_{f_1}, s_{f_2}, s_{f_3},  s_{n_1}\}$. 
		
	\end{example}


	In IC-NOMA, the index coded packets of far users are combined with index coded packets of near users suitably at two power levels as in a conventional NOMA using SC. The near users retrieve the index codes designed for them through SIC, followed by the decoding of desired data. Hence in Example \ref{ICNOMAexp2}, the total number of required transmissions in IC-NOMA to serve $N = 5$ users is $l^{\scriptscriptstyle IC-NOMA} = 3$, whereas traditional index coding requires atleast $4$ transmissions.
	\section{Design of IC-NOMA}
	The spectral efficiency offered by the IC-NOMA system can be achieved only through the innovative design of the index codes so as to reap the advantages of NOMA through SC. In order to develop an index code design for NOMA, the proposed system groups the users into two as far users and near users, by considering the channel gain of each user from the BS. The proposed algorithm is specifically developed for situations where the users can be divided into two groups based on their respective channel gains but can be extended to a more general scenario with more number of groups. The channel gain difference between any two users in the same group is considered as very small while the difference between channel gains of two users in two different groups could be considerably greater.
	
	We search for the maximum and minimum channel gains $g_{max}$ and $g_{min}$  from a group of $N$ users. Let the index set of far users be denoted as $\mathcal{I}_f$ and  the index set of near users as $\mathcal{I}_n$, where $\mathcal{I}_f = \{i:\quad |g_{max}-g_{i}|>|g_{min}-g_{i}|\}$ and $\mathcal{I}_n = \{i:\quad |g_{max}-g_{i}|<|g_{min}-g_{i}|\}$, where $i \in\ [N]$. 
	
	Consider the scenario where there are n messages. Let $x = [x_1 x_2 ...x_n]^T $ be the input vector.
	After grouping far and near users separately, we design an index code of length $l_f$ for far users based on the elements in known set $\mathcal{K}_{f} = \{\mathcal{K}_{i}:  \quad i\in \mathcal{I}_{f}\}$ and elements in want set $\mathcal{W}_{f}=\{ \mathcal{W}_{i}:  \quad i\in \mathcal{I}_{f}\} $. Let $\textbf {y}_f$ denote the index code designed for far users with encoding matrix as $\textbf {L}_f$. Then $\textbf{y}_f=\textbf {L}_f\textbf {x}$.
	
	After designing index codes for far users, we design an index code $\textbf{y}_n^c$ of length $l_n$ for the near users. Since every near user decodes the data of far users to perform SIC according to NOMA principles, the known set of each near user is updated based on the set of coded side information denoted by $\mathcal{Y}_f$ as $\mathcal{K}_{i}'= \{\mathcal{K}_{i}\cup \mathcal{Y}_f: \quad i\in \mathcal{I}_{n}\}$ . The index codes for near users are designed based on $\mathcal{K}_{n}' = \{\mathcal{K}_{i}':  \quad i\in \mathcal{I}_{n}\} $ and  $\mathcal{W}_{n}'=\{ \mathcal{W}_{i}\backslash \mathcal{W}_i^c:  \quad i\in \mathcal{I}_{n}\} $, where $\mathcal{W}_i^c$ represents the set of demands of near user $V_i$ satisfied by the index code of far users. Then $\textbf{y}_n^c=\textbf {L}_n^c\textbf {x}$. \\	
	Hence the conventional IC problem is converted to two IC problems in IC-NOMA. One is a conventional index coding problem for far users, and other is an ICCSI problem for near users. The index code design for IC-NOMA is given by Algorithm \ref{alg1}.
	\begin{algorithm}[]
	\caption{Grouping of users and updating near users}
	\begin{algorithmic}[1]
		\REQUIRE $g_i$, $\mathcal{K}_{i}$ and $\mathcal{W}_{i}$ $\forall i\in [N]$, $x = [x_1 \hspace{0.1cm} x_2 ...x_n]^T $, $N$.  
		
		\STATE $g_{max}=max_{i\in[N]}(g_i)$.
		\STATE $g_{min}=min_{i\in[N]}(g_i)$.
		\STATE $\mathcal{I}_f= [N]$.
		
		$\mathcal{I}_n= [N]$.
		
		\FOR{i=1 to $N$}     
		
		\IF {$|g_{max}-g_i|>|g_{min}-g_i|$}
		\STATE $\mathcal{I}_n=\mathcal{I}_n\backslash\{i\}$.
		\ELSE
		
		\STATE $\mathcal{I}_f=\mathcal{I}_f \backslash \{i\}$.
		
		\ENDIF
		
		\ENDFOR
		\STATE $N_f= |\mathcal{I}_f|$,
		$N_n= |\mathcal{I}_n|$.
		\STATE $\mathcal{K}_{f} = \{\mathcal{K}_{i}:  \quad i\in \mathcal{I}_{f}\}$. 
		\STATE $\mathcal{W}_{f}=\{ \mathcal{W}_{i}:  \quad i\in \mathcal{I}_{f}\} $.
		\STATE  
		Design encoding matrix 
		$\textbf{L}_f =\mathcal{F}(n, N_f,\mathcal{K}_{f}, \mathcal{W}_{f})$.
		\STATE $\textbf{y}_f=\textbf{L}_f\textbf{x}$.
		\STATE Let $\mathcal{Y}_f$ denote the set of index coded packets designed for far users.
		\STATE  $\mathcal{Y}_f=\{y_{f_{i}}:\quad i\in [l_f]\}$.
		
		\STATE $\mathcal{K}_{i}' = \{\mathcal{K}_{i}\cup \mathcal{Y}^f:  \quad i\in \mathcal{I}_{n}\}$.
		\STATE $\mathcal{K}_{n}' = \{\mathcal{K}_{i}':  \quad i\in \mathcal{I}_{n}\} $.
		\STATE Let $\mathcal{W}_i^c$ denotes the set of demands of near user $V_i$ satified by coded side information of far users.
		\STATE $\mathcal{W}_{n}'=\{ \mathcal{W}_{i}\backslash \mathcal{W}_i^c:  \quad i\in \mathcal{I}_{n}\} $.
		
		\STATE 	Design encoding matrix $\textbf{L}_n^c=\mathcal{F}(n, N_n, \mathcal{K}_{n}', \mathcal{W}_{n}')$.
		\STATE $\textbf{y}_n^c=\textbf{L}_n^c\textbf{x} $.
		\RETURN $\textbf{y}_f, \textbf{y}_n^c$.
		
	\end{algorithmic}
	\label{alg1}
\end{algorithm}

	IC-NOMA includes a set of NOMA messages superposing the index coded packets and a set of index coded messages, if needed to satisfy the user needs. Each NOMA transmission of IC-NOMA superposes two index coded packets in the power domain and is similar to a conventional NOMA with two power levels. The transmitted set of NOMA messages and the set of index code messages in the IC-NOMA system are denoted respectively as

	
	\vspace{-0.5cm}
	\begin{align}
	S_{\scriptscriptstyle k}^{\scriptscriptstyle IC- NOMA}&= 
	\begin{cases}\vspace{0.3cm}
	S_{\scriptscriptstyle k}^{\scriptscriptstyle NOMA} \hspace{1cm}
	k\in [l^{\scriptscriptstyle NOMA}]\\\vspace{0.3cm}
	\&\\\vspace{0.3cm}
	S_{\scriptscriptstyle k}^{\scriptscriptstyle IC} \hspace{1cm}
	k= \{l^{\scriptscriptstyle NOMA}+1,...l^{\scriptscriptstyle IC-NOMA}\},\\
	\end{cases}
	\end{align}		
	
	\vspace{-0.1cm}		
	\begin{align}
	S_{\scriptscriptstyle k}^{\scriptscriptstyle NOMA}&= \sqrt{\alpha P} s_{n_k} + \sqrt{(1-\alpha)P} s_{f_k}, \hspace{0.5cm}
	k \in [l^{\scriptscriptstyle NOMA}]\nonumber\\\vspace{0.5cm}
	S_{\scriptscriptstyle k}^{\scriptscriptstyle IC}&=
	\begin{cases}
	\sqrt{P} s_{f_k}, \hspace{0.3cm} k=\{l^{\scriptscriptstyle NOMA}+1,...l^{\scriptscriptstyle IC-NOMA}\},\\ \hspace{4cm} \text{ if  $l_f> l_n$ } \\
	\text{or}\\
	\sqrt{P} s_{n_k}, \hspace{0.3cm} k =\{l^{\scriptscriptstyle NOMA}+1,...l^{\scriptscriptstyle IC-NOMA}\},\\ \hspace{4cm}\text{if  $l_f<l_n$},
	\end{cases}
	\end{align}	
	
\noindent	where $s_{f_i}=enc(y_{f_i})$, $i\in [l_f]$ and $s_{n_i}=enc(y_{n_i}^c)$, $i\in [l_n]$. The required number of NOMA transmissions in IC-NOMA system is $l^{\scriptscriptstyle NOMA}=min(l_f,l_n)$. If $l^{\scriptscriptstyle IC-NOMA}$ represents the total number of transmissions needed in IC-NOMA system as NOMA and IC, then $l^{\scriptscriptstyle IC-NOMA}=max(l_f,l_n)$. 
	\vspace{0.3cm}
	
	Hence the IC-NOMA system consists of three different cases as follows:
	\vspace{0.2cm}\\
	$ Case \hspace{0.2cm}I:$ When $l_f=l_n=l_{\scriptscriptstyle NOMA}$, demands of both the far and near users are satisfied by $l_{\scriptscriptstyle NOMA}$ number of NOMA transmissions. \vspace{0.2cm}\\ 
	$ Case\hspace{0.2cm} II:$ When $l_f > l_n$, a total of $l_f$ transmissions will be required by IC-NOMA system, with $l_n$ number of NOMA transmissions and  $l_f-l_n$ number of index coded transmissions.  Far user receives NOMA transmissions and index coded transmissions to decode the desired data. Near user decodes far user data from NOMA and index coded transmissions (as coded side information) and perform SIC to decode the desired data.
	\vspace{0.2cm}\\ 
	$ Case \hspace{0.2cm}III:$ When $l_f < l_n$, a total of $l_n$ transmissions will be required by IC-NOMA system, with $l_f$ number of NOMA transmissions and  $l_n-l_f$ number of index coded transmissions.  Far users requires only NOMA transmissions to decode the desired data. Near users requires NOMA and index coded transmissions to decode the desired data.

	\begin{claim}For a given index coding problem, IC-NOMA serves the user demands with less than or equal number of transmissions compared to conventional IC.
		\begin{proof}
			For a given index coding problem with $N$ number of users, let $l^{\scriptscriptstyle IC}$ be the optimal length of a conventional index code.
			Considering the same problem as two sub-problems in the IC-NOMA scenario, let $l_f$ be the length of index code designed for far users with $N_f < N$ and $l_n$ be the length of index codes designed for near users with $N_n < N$.
			Since the conventional index coding problem is split as two subproblems in IC-NOMA, $l_f \le l^{\scriptscriptstyle IC}$ and $l_n \le l^{\scriptscriptstyle IC}$, and as given above $l^{\scriptscriptstyle IC-NOMA} = max(l_f,l_n)$. Hence $l^{\scriptscriptstyle IC-NOMA} \le l^{\scriptscriptstyle IC}$. Thus IC-NOMA serves the user demands with less than or equal number of transmissions compared to conventional IC.
		\end{proof}
	\end{claim}
	
	
	\section{Rate Analysis}
	\label{rate}
	Let $N_f$ be the number of far users with channel gain of each user as $g_i \approxeq g_f$, for $i \in\mathcal{I}_f$, where  $\mathcal{I}_f$ denotes the index set of far users. Let $N_n$ be the number of near users with channel gain of each user as $g_i \approxeq g_n$, for $i \in \mathcal{I}_n$, where  $\mathcal{I}_n$ denotes the index set of near users. We have $g_n > g_f$. The achievable rate at each user for a given transmission depends on their respective channel conditions.  The rate analysis is performed at each receiver by assuming additive white Gaussian noise with zero mean and unit variance. All the rates in subsequent discussions are expressed in bits per channel use.
%
	
	We quantify the improvement in achievable information rate of the proposed IC-NOMA system over conventional IC. 
	
	\subsection{Conventional IC Scenario}
 Let P be the power required per transmission in a conventional IC scenario. Let the number of index coded transmissions needed to satisfy user demands be  $l^{\scriptscriptstyle IC}$. Thus the transmitted signal is 
	\begin{equation}
	S_i^{\scriptscriptstyle IC}=\sqrt{P}enc(y^i)\hspace{0.3cm} i=\{1, 2, ....l^{\scriptscriptstyle IC}\}.
	\end{equation}
	The achievable rates at each near and far users in index coding problem is denoted as $R_f^{IC}$ and $R_n^{IC}$ and are given by 
	\begin{equation}
	R_f^{\scriptscriptstyle IC}=log_2\left(1+g_fP\right),
	\end{equation}
	\begin{equation}
	R_n^{\scriptscriptstyle IC}=log_2(1+g_nP).
	\end{equation}
	Since $g_n > g_f$, we have $R_f^{IC} <R_n^{IC}$. Thus the rate that can be supported in index coded transmission is limited by that of the far user as,  \begin{equation}
	R^{\scriptscriptstyle IC}=R_f^{\scriptscriptstyle IC}.
	\label{R1}
	\end{equation}

	\subsection{IC-NOMA Scenario}
	
	In proposed IC-NOMA scheme, the data transmitted could be either simple index coded data or the index coded data combined with NOMA, as seen earlier. In proposed IC-NOMA scheme, the achievable information rate for NOMA transmissions is indicated as $R^{\scriptscriptstyle IN-NOMA}$. The achievable information rate for index coded transmissions of IC-NOMA system in Case II and Case III are represented as $R^{\scriptscriptstyle IN-IC^{(i)}}$ with $i= 2$ and $i=3$ respectively. 
	\subsubsection{CASE I  ($l_f=l_n $)}
	\vspace{0.2cm}
	In this case IC-NOMA system will have NOMA transmissions only to satisfy user demands. For comparison, the power per transmission is considered to be the same as that of conventional index coded transmission ie., $ P$. Then,
	\begin{align}
	S_{\scriptscriptstyle k}^{\scriptscriptstyle IC-NOMA}&=
	S_{\scriptscriptstyle k}^{\scriptscriptstyle NOMA}\nonumber\\
	&= \sqrt{\alpha P} s_{n_k} +\sqrt{(1-\alpha)P} s_{f_k} \hspace{0.2cm}	k\in [l_{\scriptscriptstyle NOMA}].\nonumber\\
	\end{align} \\
	\vspace{0.3cm}
	Let $R_{f}^{\scriptscriptstyle IN-NOMA}$ denote the achievable rate at far user decoding the data from NOMA transmission. Then, \begin{equation}
	R_{f}^{\scriptscriptstyle IN-NOMA}=log_2\left(1 + \frac{ (1-\alpha)P g_f}{\alpha P g_f + 1}\right).
	\end{equation}
	Near user performs SIC of far user data.
	Let $R_{n}^{\scriptscriptstyle IN-NOMA}$ denote the achievable rate at near user. Then, \begin{equation}
	R_{n}^{\scriptscriptstyle IN-NOMA}=log_2(1 + \alpha Pg_n).
	\end{equation}
	Hence sum rate of transmission channel for this case denoted as $R^{\scriptscriptstyle IN-NOMA}$ for each NOMA transmission of IC-NOMA system is  \begin{align}
	R^{\scriptscriptstyle IN-NOMA}&=R_f^{\scriptscriptstyle IN-NOMA} + R_n^{\scriptscriptstyle IN-NOMA}.\nonumber\\
	&=log_2\left\{ \frac{1 + Pg_f}{1+\alpha Pg_f} (1 +\alpha Pg_n)\right\}.
	\label{R2}
	\end{align}
	Consider \eqref{R1} and \eqref{R2} to analyze the the improvement in the rate that can be supported by the proposed system.
	\begin{align}
	R^{\scriptscriptstyle IN-NOMA}-R^{\scriptscriptstyle IC}
	&=log_2\left(\frac{1+\alpha P g_n}{1+\alpha P g_f}\right).
		\label{tra1}
	\end{align}
	Since $g_n > g_f$, \begin{align}
&R^{\scriptscriptstyle IN-NOMA} > R^{\scriptscriptstyle  IC}.
\end{align}	
Hence the proposed system can support a higher information rate compared to conventional IC.
	\subsubsection{CASE II ($l_f > l_n$)}
	In this case, there will be a total of  $l_f$ number of transmissions in IC-NOMA, with $l_n$ NOMA transmissions and $l_f-l_n$ index coded transmissions. The IC-NOMA transmissions are given by 
	\begin{align}
	S_{\scriptscriptstyle k}^{\scriptscriptstyle IC-NOMA}= 
	\begin{cases}\vspace{0.3cm}
	S_{\scriptscriptstyle k}^{\scriptscriptstyle NOMA} =\sqrt{\alpha P} s_{n_k} +\sqrt{(1-\alpha)P} s_{f_k},\\ \nonumber \hspace{4cm} k\in [l_{n}]\nonumber\\
	\vspace{0.3cm}
	\text{and} \\
	S_{\scriptscriptstyle k}^{\scriptscriptstyle IC} =\sqrt{P} s_{f_k}, \hspace{0.2cm} k =\{l_{n}+1,...l_f\}.
	\end{cases}\\
	\end{align}
	The far user receives NOMA transmissions and IC transmissions to decode the desired data. Near user decodes far user data from NOMA and IC transmissions (as coded side information) to perform SIC to decode the desired.
	
	The information rate for NOMA transmission of IC-NOMA will be the same as \eqref{R2}. The index coded transmissions in this case are index coded packets designed for far users. The information rate that can be supported by IC-NOMA under this scenario is decided by the information rate supported by the IC transmission of IC-NOMA, which is nothing but the achievable data rate at far user. Therefore achievable rate of IC-NOMA for IC transmission for Case II is calculated as\begin{equation}
	R^{\scriptscriptstyle IN-IC^{(2)}}=log_2(1 + Pg_f) = R^{\scriptscriptstyle  IC}.
	\label{R3}
	\end{equation}	
	 The improved information rate that can be supported by NOMA transmissions is given in \eqref{tra1}. From \eqref{R3}, it can be seen that, the achievable rate of index coded transmission is same as that of conventional IC.

	\subsubsection{CASE III  ($l_f < l_n$)}
	Under this scenario, there will be a total of  $l_n$ number of transmissions in IC-NOMA, with $l_f$ NOMA transmissions and $l_n-l_f$ IC transmissions. The IC-NOMA transmissions are given by 
	
	\begin{align}
	S_{\scriptscriptstyle k}^{\scriptscriptstyle IC-NOMA}= 
	\begin{cases}\vspace{0.3cm}
	S_{\scriptscriptstyle k}^{\scriptscriptstyle NOMA} =\sqrt{\alpha P} s_{n_k} +\sqrt{(1-\alpha)P} s_{f_k},\\\nonumber \hspace{4cm} k\in [l_{f}]\nonumber\\
	\vspace{0.3cm}
	\text{and} \\
	S_{\scriptscriptstyle k}^{\scriptscriptstyle IC} =\sqrt{P} s_{n_k}, \hspace{0.2cm} k=\{l_{f}+1,...l_n\}.
	\end{cases}\\
	\end{align}
	Far users require only NOMA transmissions to decode the desired data.
	Near users require NOMA and IC transmissions to decode the desired data. Near users decode far user data from NOMA transmissions (as coded side information) to perform SIC to decode desired.
	The achievable rate for NOMA transmission of IC-NOMA is given by \eqref{R2}.\\
	The achievable rate for IC transmission of IC-NOMA system under Case III is that of the near user data rate and is given as, \begin{equation}
	R^{\scriptscriptstyle IN-IC^{(3)}}=log_2(1 + Pg_n).
	\label{R4}
	\end{equation}
	The improvement in achievable rate for index coded transmission of IC-NOMA compared to conventional IC can be obtained from \eqref{R1} and \eqref{R4} as,  
	\begin{align}
	R^{\scriptscriptstyle IN-IC^{(3)}}-R^{\scriptscriptstyle IC}
	&=log_2\left(\frac{1+ P g_n}{1+ P g_f}\right).
	\end{align}
The improvement in achievable information rate for NOMA transmissions is given in \eqref{tra1}. It can be seen that there is an improvement in the achievable information rate for both IC and NOMA transmissions of IC-NOMA system in this case.

Thus it can be seen that the achievable rate for proposed IC-NOMA is greater than conventional IC under Case I and Case III for all transmissions. For Case II the achievable rate for NOMA transmission is greater than conventional IC while the rate for IC transmission is same as in conventional IC.

	\section{TRANSMISSION POWER ANALYSIS}
	\label{ppower}
	In this section, we demonstrate how the proposed IC-NOMA system outperforms traditional IC in terms of power consumption. The power per transmission required for the IC-NOMA system to provide an achievable information rate at least as good as that of conventional IC is evaluated. 
	
	Let $P^{\scriptscriptstyle IC}$ be the power per transmission for conventional IC, $P_a$ be the power per NOMA transmission for IC-NOMA and $P_b^{(i)}$ for $i =2$ and $i =3$ be the power per index coded transmission of IC-NOMA system for Case II and Case III. 
	
	The total power consumption is computed for IC and IC-NOMA scheme, to quantify the power saving of the proposed system. \vspace{-0.5cm}
	\subsection{Case I ($l_f=l_n=l^{\scriptscriptstyle NOMA}$):}
	Considering $P^{\scriptscriptstyle IC}$ as the power per transmission for conventional IC and $P_a$ as the power per NOMA transmission for IC-NOMA, \eqref{R1} \eqref{R2} are modified respectively as 
	\begin{equation}
	R^{IC}=log_2(1+g_fP^{\scriptscriptstyle IC})\hspace{0.1cm},\label{p'1}
	\end{equation}
	\begin{align}
	R^{\scriptscriptstyle IN-NOMA} =log_2\left\{ \frac{1 + P_ag_f}{1+\alpha P_ag_f} (1 +\alpha P_ag_n)\right\}.
	\label{R2'}
	\end{align}
	To find the minimum power requirement for IC-NOMA transmission that makes $R^{\scriptscriptstyle IN-NOMA}$ atleast as good as $R^{\scriptscriptstyle IC}$; consider \eqref{p'1} and \eqref{R2'}.
	\begin{equation}
	log_2(1+g_fP^{\scriptscriptstyle IC}) =log_2\left\{\frac{1 + P_ag_f}{1+\alpha P_ag_f} (1 +\alpha P_ag_n)\right\} \label{r5}
	\end{equation}
Let $P^{\scriptscriptstyle IC}=P_a+\zeta$ where $-p \leq \zeta \leq +p$,\hspace{0.3cm} $p=|P^{\scriptscriptstyle IC}-P_a|$. Then,
	\begin{align}
	\zeta &=\frac{(1+P_ag_f) (\alpha P_a(g_n-g_f))}{g_f (1+\alpha P_ag_f)}.
	\label{R6}
	\end{align}
From \eqref{R6}, we have $\zeta>0 \implies P^{\scriptscriptstyle IC} > P_a$ i.e., power per transmission for IC is greater than that of IC-NOMA. Then the power per NOMA transmission of IC-NOMA system is 
	\begin{equation}
	P_a=P^{IC}-\zeta.
	\label{pnew}
	\end{equation}
	Hence we have power saving per transmission for the IC-NOMA system; now we have to find the total power saving across the system considering the required number of transmissions.\\
	Let $l^{\scriptscriptstyle IC}$ be the required number of typical IC transmissions needed to satisfy the user demands.  The total power saving denoted as $P_{\scriptscriptstyle Saving}^{(1)}$ for Case I is given by
	
	\begin{align}
	P_{\scriptscriptstyle Saving}^{(1)}&= l^{\scriptscriptstyle IC}P^{\scriptscriptstyle IC}-l^{\scriptscriptstyle NOMA}P_a. \nonumber \\
	&=(l^{\scriptscriptstyle IC}-l^{\scriptscriptstyle NOMA})P_a+ \zeta l^{\scriptscriptstyle IC} .
	\label{p1}
	\end{align}
	
	\subsection{Case II ($l_f>l_n$):}
	In this scenario, we need to calculate the power per transmission of the IC-NOMA scheme to ensure that the achievable information rate for NOMA and IC transmissions is at least as good as conventional IC.
	
	From \eqref{pnew} we know that the NOMA transmissions of IC-NOMA demands less power per transmission to support an information rate at least as good as conventional IC. Let $P_b^{(2)}$ be the power per IC transmission for Case II of IC-NOMA system. Then \eqref{R3} is modified as 
	\begin{equation}
	R^{\scriptscriptstyle IN-IC^{(2)}}=log_2(1 + P_b^{(2)}g_f).
	\label{R3new}
	\end{equation}	
From \eqref{p'1} and \eqref{R3new}, it can be seen that the power per transmission required will be the same for both IC transmission of IC-NOMA and conventional IC to have equal achievable rate. Then,
\begin{equation}
	P_b^{(2)}=P^{\scriptscriptstyle IC}.
\end{equation}
Hence in the IC-NOMA system with $l_f>l_n$, we have non-zero power saving for all the $l_n$ number of NOMA transmissions and zero power savings for $l_f-l_n$ number of IC transmissions; when compared to conventional IC.
	In this case the total power saving denoted as $P_{\scriptscriptstyle Saving}^{(2)}$ for Case II is given by\\
	\begin{align}
	P_{\scriptscriptstyle Saving}^{(2)} &= l^{\scriptscriptstyle IC}P^{\scriptscriptstyle IC}-(l_{n}P_a+(l_f-l_n) P^{\scriptscriptstyle IC})\nonumber \\
	&=(l^{\scriptscriptstyle IC}-l_{f})P^{\scriptscriptstyle IC} + \zeta l_n.
	\label{p2} 
	\end{align}
	\subsection{Case III ($l_f < l_n$):}
	The NOMA transmission of the IC-NOMA system saves power compared to conventional IC to achieve equal information rate as given by \eqref{pnew}. 
	
	Considering $P_b^{(3)}$ as the power per IC transmission for Case III of IC-NOMA system,  \eqref{R4} is modified as
	\begin{equation}
	R^{\scriptscriptstyle IN-IC^{(3)}}=log_2(1 + P_b^{(3)}g_n).
	\label{R4new}
	\end{equation}
	To find the minimum power per transmission required to make the information rate for the IC transmission of IC-NOMA system atleast as good as that of conventional IC, consider \eqref{p'1} and \eqref{R4new}.
	\begin{equation}
	log_2(1+P^{\scriptscriptstyle IC}g_f)=log_2(1+P_b^{(3)}g_n) .\\
	\label{l1}
	\end{equation}
	Let $P^{\scriptscriptstyle IC}=P_b^{(3)}+\zeta_1$, where $-p_1 <\zeta_1 <p_1$, $p_1=|P^{\scriptscriptstyle IC}-P_b^{(3)}|$. Then, 
	\begin{align}
	&\zeta_1=\frac{(g_n-g_f)P_b^{(3)} }{g_f}.
	\label{R7}
	\end{align}
	From \eqref{R7}, $\zeta_1>0 \implies P^{\scriptscriptstyle IC} > P_b^{(3)}$, i.e., power per transmission for conventional IC is greater than that of IC transmission of IC-NOMA. The power per transmission required for IC transmissions of IC-NOMA system to provide an information rate atleast as good as IC is 
	
	\begin{equation}
		P_b^{(3)}=P^{\scriptscriptstyle IC}-\zeta_1.
	\end{equation}
	
\noindent	Thus for $l_f < l_n$, we have power savings in the IC-NOMA system for NOMA and IC transmissions.
	\\To find the total power saving across the system, consider the required number of transmissions\\
	\begin{align}
	P_{\scriptscriptstyle Saving}^{(3)} &= l^{\scriptscriptstyle IC}P^{\scriptscriptstyle IC}-(l_{f}(P^{\scriptscriptstyle IC}-\zeta) +(l_n-l_f)(P^{\scriptscriptstyle IC}-\zeta_1)). \nonumber \\
	&=(l^{\scriptscriptstyle IC}-l_n)P^{\scriptscriptstyle IC}+l_f\zeta+(l_n-l_f)\zeta_1.
	\label{p3} 
	\end{align}
	From \eqref{p1}, \eqref{p2}, \eqref{p3}, it can be seen that for providing same information rate, the IC-NOMA system will have positive power savings when compared to conventional IC.
	
	\section{QUALITY OF SERVICE (QoS) REQUIREMENT ANALYSIS}
	\label{qqos}
		This section compares the performance of the proposed IC-NOMA system with that of the conventional IC in terms of transmission power for meeting the rate required to satisfy quality of service demands of the user. The following discussions assume R as the rate requirement  at each individual user to meet QoS. The minimum transmission power required to meet the QoS requirement under both schemes  are analyzed here.
		
		Let $P^{IC}$ be the power per transmission required for conventional IC to meet the QoS requirement. In IC-NOMA to  meet the QoS requirement, consider the power per NOMA transmission as $P_c$ and power per IC transmission for Case II and Case III as $P_d^{(2)}$ and $P_d^{(3)}$ respectively.  
	\subsection{Conventional IC Scenario}
	Let $P^{\scriptscriptstyle IC}$ be the power per transmission for conventional IC. The achievable rates at near and far users due to IC transmission is as follows:
	\begin{align}
	R_f^{\scriptscriptstyle IC}&=log(1+P^{\scriptscriptstyle IC}g_f),\\
	R_n^{\scriptscriptstyle IC}&=log(1+P^{\scriptscriptstyle IC}g_n).
	\end{align}
	The power per transmission $P^{\scriptscriptstyle IC}$ required to meet QoS requirement R is given by:
	\begin{align}
	&P^{\scriptscriptstyle IC}=max\left(\frac{2^R-1}{g_n},\frac{2^R-1}{g_f}\right)\\
	&=\frac{2^R-1}{g_f}.
	\label{pp1}		
	\end{align}
	Let $l^{\scriptscriptstyle IC}$ be the required number of transmissions in conventional IC scenario. The total power requirement for each transmission to meet the QoS requirement is
	\begin{align}
	P_{total}^{\scriptscriptstyle IC} =\frac{2^R-1}{g_f} \hspace{0.1cm} l^{\scriptscriptstyle IC}=P^{\scriptscriptstyle IC} \hspace{0.1cm}l^{\scriptscriptstyle IC}.\label{pp111}
	\end{align}
	\subsection{IC-NOMA Scenario}
	\subsubsection{Case I ($l_f=l_n=l^{\scriptscriptstyle NOMA}$)} 
	The far and near users satisfy their demands using $l^{\scriptscriptstyle NOMA}$ number of NOMA transmission only.
	The achievable rates at near and far users are given respectively as
	\begin{align}
	&R_n^{\scriptscriptstyle IN-NOMA}=log(1+\alpha P_cg_n),\\
	&R_f^{\scriptscriptstyle IN-NOMA}=log\left(1+\frac{(1-\alpha) P_cg_f}{\alpha P_cg_f +1}\right).\label{pp11}
	\end{align}
Let $P_{c_{n}}$ and $P_{c_{f}}$ be the minimum transmission powers required to meet the QoS requirement $R$ at near and far users, respectively. Then, 
	\begin{align}
	P_{c_{n}}&=\frac{2^R-1}{\alpha g_n}=\frac{P^{\scriptscriptstyle IC}g_f}{\alpha g_n},
	\end{align}
	\begin{align}
		P_{c_{f}}&= \frac{2^R-1}{g_f (1-\alpha-\alpha (2^R-1))}\nonumber\\
	&= \frac{P^{\scriptscriptstyle IC}}{1-\alpha-g_f\alpha P^{\scriptscriptstyle IC}}.
	\end{align}
Hence the transmission power $P_c$ required to meet QoS requirmenent R for both far and near users is
	\begin{align}
	P_c	&=max\left(\frac{P^{\scriptscriptstyle IC}g_f}{\alpha g_n}, \frac{P^{\scriptscriptstyle IC}}{1-\alpha-g_f\alpha P^{\scriptscriptstyle IC}}\right)	.
	\label{pp22}
	\end{align}
	In this case the total power requirement of IC-NOMA system with $l^{\scriptscriptstyle NOMA}$ number of NOMA transmissions is given by
	\begin{align}
	P_{total}^{\scriptscriptstyle IC-NOMA^{(1)}} &=P_c \hspace{0.1cm}l^{\scriptscriptstyle NOMA}.\label{pp2}
	\end{align}
	The power per NOMA transmission $P_c$ will be always greater than $P^{\scriptscriptstyle IC}$ due to superposition coding. 
    Now whether IC-NOMA can provide power saving compared to IC depends upon how much  $l^{\scriptscriptstyle NOMA}$ is smaller than $l^{\scriptscriptstyle IC}$. It can be seen that for the case $l^{\scriptscriptstyle NOMA}= l^{\scriptscriptstyle IC}$, IC-NOMA requires more power than conventional IC. 
	
	\subsubsection{CASE II ($l_f >l_n$)}
	The far and near users receive NOMA and index coded transmissions to meet their demands. There will a total of  $l_f$ number of transmissions in IC-NOMA system, with $l_n$  number of NOMA transmissions and  $l_{f}-l_n$ number of IC transmissions. 
	The transmission power required for IC transmission of IC-NOMA to meet QoS is given as \begin{align}P_d^{(2)}=P^{\scriptscriptstyle IC}=\frac{2^R-1}{g_f} .\end{align}
	In this case, the total power requirement of the IC-NOMA system to meet the QoS requirement at each transmission is given by
	\begin{align}
	P_{total}^{\scriptscriptstyle IC-NOMA^{(2)}}&=P_c\hspace{0.1cm}	l_n +  P_d^{(2)}\hspace{0.1cm}(l_f - l_n)\nonumber\\
	&= P_d^{(2)} \hspace{0.1cm}l_f + (P_c-P_d^{(2)})\hspace{0.1cm}l_n.
	\label{pp3}
	\end{align}
Here also power saving of IC-NOMA depends on the values of  $l_f$  and $l_n$ . There could be power saving when $l_f$  is less than $l^{\scriptscriptstyle IC}$ . When $l_f= l^{\scriptscriptstyle IC}$ and $l_n = 0$, the power required by IC-NOMA system will be same as that required by conventional IC. But when  $l_f= l^{\scriptscriptstyle IC}$ and $l_n \ne 0$, the total power required by IC-NOMA system will be more than conventional IC.
	\subsubsection{CASE III ($l_f < l_n$)}
	This scenario of the IC-NOMA system consists of both NOMA and index coded transmissions. The far users satisfy their respective demands from NOMA transmissions, whereas near users require both NOMA and index coded transmissions to satisfy the respective demands.
	In this case,  IC-NOMA system will have a total of $l_n$ transmissions with $l_f$  NOMA transmissions and  $l_{n}-l_f$ number of index coded transmissions.\\
	The transmission power required for IC transmission of IC-NOMA to meet QoS is given as \begin{align}P_d^{(3)}=\frac{2^R-1}{g_n} = P^{\scriptscriptstyle IC}\hspace{0.1cm}\frac{g_f}{g_n}.\end{align}
Since $\frac{g_f}{g_n} <1$, the power required per IC transmission of IC-NOMA system is less than that required by conventional IC to meet the QoS.
 The total power consumption is calculated as
	\begin{align}
	P_{total}^{\scriptscriptstyle IC-NOMA^{(3)}}&=P_c\hspace{0.1cm}l_f + P_d^{(3)} \hspace{0.1cm}(l_{n}-l_f)\nonumber\\
	&=(P_c-P_d^{(3)})\hspace{0.1cm}l_f + P_d^{(3)} \hspace{0.1cm}l_n.
	\label{pp4}
	\end{align}
	In this case also, the power saving of IC-NOMA compared to IC depends upon the individual values of $l_f$ and $l_n$. But in this case there could be power saving even when  $l_n =l^{IC}$, since $P_d^{(3)} < P^{\scriptscriptstyle IC}$. 	
	\section{SELECTION OF INDEX CODE FOR THE FAR USERS}
The discussions on transmit power requirement for the QoS in Section \ref{qqos}, throws light on to the fact that  the proposed algorithm for IC-NOMA can be  modified further to improve the power efficiency. From the power efficiency calculations, the following  two conclusions can be drawn:
\begin{itemize}
	\item  For the cases $l^{\scriptscriptstyle IC} =l_f =l_n$  and $l^{\scriptscriptstyle IC} = l_f >l_n \ne 0$, the IC-NOMA system demands more power than conventional IC. In this case, it is beneficial to proceed with conventional IC. 
	\item For all other cases, the power saving of IC-NOMA is directly depending upon the individual values of $l_f$ and $l_n$. 
\end{itemize} 
Moreover, for a given  $l_f$  the scheme with minimum value for $l_n$  can minimize the overall power consumption. Considering this fact, it is required to arrive at an index code design that minimizes $l_n$ for a given $l_f$. For a given $l_f$ there can be many possibilities for far user index code design. Since actual design of far user index code decides the coded side information of the near user, the length $l_n$ of the near user index code depends heavily on the actual design of index code for far user. This can be made clear by considering Example \ref{exmp3}.
\begin{example}
	\label{exmp3}
	Consider an IC problem with $N=5$, let the known set distribtion be the same as in Table \ref{ICNOMAexp3} and the want set distribution be as in Table \ref{spc}. Table \ref{spccodes} shows the set of possible index codes designed for scenario under consideration. It can be seen that the optimal length of conventional index code for this problem is $l^{\scriptscriptstyle IC}=4$.

	Table \ref{spccodes} shows the variation in length of near user index code according to the change index code design for far users. From Table \ref{spccodes} it can be seen that the length of index code for near user $l_n$ varies from 3 to 1 for the same value of $l_f$.

	\begin{table}[h]
	\centering
		\caption{The want set distribution for $N=5$ with the known set same as given in Table \ref{ICNOMAexp3}.}
	\begin{tabular}{|c|c|c|c|}
		\hline
		User	& Want set \\
		\hline	
		$V_i$	&  $\mathcal{W}_i$\\
		\hline
		$V_1$	&  $\{ x_4, x_5, x_7\}$   \\
		\hline
		$V_2$	& $\{x_5, x_6\}$  \\
		\hline
		$V_3$	&   $\{ x_1, x_6\}$  \\
		\hline
		$V_4$	&  $\{x_1, x_2, x_3\}$  \\
		\hline
		$V_5$	&  $\{x_1\}$ \\
		\hline
	\end{tabular}

	\label{spc}
\end{table}

\begin{table}[H]
	\tabcolsep=0.04cm
		\centering
			\caption{The set of possible index code designs for IC-NOMA system by considering want set distribution as in Table \ref{spc} and known set as in Table \ref{ICNOMAexp3}. }
		\begin{tabular}{|c|c|c|}
			\hline
			  &\multicolumn{2}{|c|}{Index codes designed.}\\
			\hline	
	$l_f \hspace{0.1cm} \& \hspace{0.1cm} l_n$	& 	$\mathcal{Y}_f$& $\mathcal{Y}_n^c$ \\
			\hline
$l_f=3$, $l_n =3$&	$\{	x_1, x_2+x_7, x_3\}$	&  $\{ x_2 +x_5, x_3 +x_6, x_4 +x_1\}$  \\\hline
$l_f=3$, $l_n =2$ &$\{x_1 +x_6, x_2 +x_7,x_3 +x_6 \}$ &$\{x_2 +x_5, x_4\}$
\\\hline
$l_f=3$, $l_n =1$ &$\{x_1 +x_7, x_2 +x_5,x_3 +x_6\}$ &$\{x_4 +x_1\}$
\\\hline
		\end{tabular}
	
		\label{spccodes}
	\end{table}

\end{example}


 Hence there is a need to design index code for far users by considering the index coding problem of near users, such that the solution to the far user index coding problem minimizes the index code length of the near users. Thus the proposed algorithm for IC-NOMA is modified by considering the above insights. The modified algorithm is presented as Algorithm \ref{alg2}.
 
 Consider $l_f$ as the optimal index code length corresponding to the index coding problem of far users. Let $\mathcal{L}_f$ be the set of all possible optimal index codes designed for far users. The index code $\textbf{L}_{f}^j \in \mathcal{L}_f$ with $j \in \left[|\mathcal{L}_f|\right] $ that minimize $|\textbf{y}_{n}^{c^j}|$ can give the best performance in terms of power efficiency. Algorithm \ref{alg2} gives the modified design of index code for IC-NOMA.

	\begin{algorithm}[H]
	\caption{Selection of index code for far users}
	\begin{algorithmic}[1]
		\REQUIRE $g_i$, $\mathcal{K}_{i}$ and $\mathcal{W}_{i}$ $\forall i\in [N]$, $x = [x_1 \hspace{0.1cm} x_2 ...x_n]^T $, $N$.  
		
		\STATE $g_{max}=max_{i\in[N]}(g_i)$.
		\STATE $g_{min}=min_{i\in[N]}(g_i)$.
		\STATE $\mathcal{I}_f= [N]$.
		
		$\mathcal{I}_n= [N]$.
		
		\FOR{i=1 to $N$}     
		
		\IF {$|g_{max}-g_i|>|g_{min}-g_i|$}
		\STATE $\mathcal{I}_n=\mathcal{I}_n\backslash\{i\}$.
		\ELSE
		
		\STATE $\mathcal{I}_f=\mathcal{I}_f \backslash \{i\}$.
		
		\ENDIF
		
		\ENDFOR
		\STATE $N_f= |\mathcal{I}_f|$,
		$N_n= |\mathcal{I}_n|$.
		\STATE $\mathcal{K}_{f} = \{\mathcal{K}_{i}:  \quad i\in \mathcal{I}_{f}\}$. 
		\STATE Let $l_f$ be optimal index code length considering IC problem for the far users.
		\STATE  $\mathcal L_{f} =\{\textbf{L}_{f}^j: \quad \textbf{L}_f^{j}=\mathcal{F}(n, |\mathcal{I}_f|, \mathcal{K}_{f}, \mathcal{W}_{f})$ ; such that length of the code is $l_f$$\}$.
		\FOR {$j= 1:|\mathcal{L}_f|$}
		\STATE $\textbf{y}_f^{j}=\textbf{L}_f^{j}\textbf{x} $; \hspace{0.3cm}$\mathcal{Y}_{f}^j$ be the set corresponding to the vector $\textbf{y}_f^{j}$.
		\STATE $\mathcal{K}_{i}^{j} = \{\mathcal{K}_{i}\cup \mathcal{Y}_f^j:  \quad i\in \mathcal{I}_{n}\}$.
		\STATE $\mathcal{K}_{n}^{j} = \{\mathcal{K}_{i}^{j}:  \quad i\in \mathcal{I}_{n}\} $.
		\STATE Let the set of demands of near user $V_i$ satified by coded side information of far users is denoted as $\mathcal{W}_i^{c^j}$.
		\STATE $\mathcal{W}_{n}^{j}=\{ \mathcal{W}_{i}\backslash \mathcal{W}_i^{c^j}:  \quad i\in \mathcal{I}_{n}\} $.
		
		\STATE 	Design encoding matrix as $\textbf{L}_n^{c^j}=\mathcal{F}(n, |\mathcal{I}_n|, \mathcal{K}_{n}^{j}, \mathcal{W}_{n}^{j})$.
		\STATE $\textbf{y}_n^{c^j}=\textbf{L}_n^{c^j}\textbf{x} $.
		
		\ENDFOR
		\STATE Choose $\{\textbf{L}_f^{j^*}, \textbf{L}_n^{c^{j^*}}\}$ pair for which the length of   $\textbf{y}_n^{c^{j^*}}=\textbf{L}_n^{c^{j^*}}\textbf{x} $ is minimum.
		\RETURN $\textbf{y}_f^{j^*}, \textbf{y}_n^{c^{j^*}}$.
		
	\end{algorithmic}
	\label{alg2}
\end{algorithm}

\begin{remark}
By selecting the optimal index code for far users that minimizes the index code length of near users as in Algorithm \ref{alg2}, the above mentioned cases $l^{\scriptscriptstyle IC} =l_f =l_n$  and $l^{\scriptscriptstyle IC} = l_f >l_n \ne 0$, will be changed to $l^{\scriptscriptstyle IC} = l_f >l_n = 0$, which is same as conventional IC.	
\end{remark}

\vspace{-0.75cm}
	\section{ILLUSTRATION OF RATE, TRANSMISSION POWER AND QoS ANALYSIS}
	In this section, we attempt to quantify the improvement in performance of the proposed IC-NOMA system when compared with conventional IC analytically with example. The example considered for analytical studies is the VANET scenario described in Example \ref{ex:ICN2} with $N = 5$ having same known set as in  Table \ref{ICNOMAexp3}  with different demand set conditions as given in Table \ref{T2}. It can be seen that for all the three want set distributions considered in the example,  the length of the optimal linear index code for conventional IC is same given as $l^{\scriptscriptstyle IC}=4$
%
%

	\begin{table}[h]
		\tabcolsep=0.1cm
		\centering
			\caption{ Three different cases of IC-NOMA system by choosing different demand set and choosing the known set same as given in Table \ref{ICNOMAexp3}.}
		\begin{tabular}{|c|c|c|c|c|}
			\hline
			User	& \multicolumn{3}{c|}{Demand set $\mathcal{D}_i$ (Before R2V Phase)} \\
			\hline	
			$V_i$	&  $l_f=l_n$ & $l_f>l_n$ & $ l_f < l_n$ \\
			\hline
			$V_1$	& $\{x_1,x_2,x_4,x_7\}$  & $\{x_4, x_7\}$ & $\{x_4, x_5, x_6\}$ \\
			\hline
			$V_2$	&  $\{x_2,x_3, x_5\}$ & $\{x_2, x_3, x_4, x_5\}$ & $\{x_1,x_3\}$ \\
			\hline
			$V_3$	&  $\{x_2,x_3, x_4,x_5, x_6\}$  & $\{x_1, x_2, x_6\}$ &$\{x_1, x_2, x_3, x_5, x_7\}$  \\
			\hline
			$V_4$	& $\{x_1,x_5, x_6, x_7\}$  & $\{x_2, x_3,x_7\}$ & $\{x_5, x_6\}$ \\
			\hline
			$V_5$	&  $\{x_3, x_6\}$ &$\{x_1,x_6\}$  & $\{x_3, x_6,x_7\}$ \\
			\hline
		\end{tabular}
	
		\label{T2}
	\end{table}

	Table \ref{T3} shows the index code design for three want set conditions developed  through Algorithm \ref{alg2} of the proposed IC-NOMA scheme.

	
		\begin{table}[H]
	\tabcolsep=0.1cm
	\centering
	\caption{ Index codes designed for three different cases of IC-NOMA system.}
	\begin{tabular}{|c|c|c|}
		\hline
		Case & $\mathcal{Y}_f$ & $\mathcal{Y}_n^c$  \\ \hline
		$l_f=l_n$& $\{x_1 +x_7, x_3+x_6\}$& $\{x_2 +x_4, x_4+x_5\}$\\ \hline
		$l_f>l_n$& $\{x_1 + x_7, x_2+x_5, x_6+x_3\}$ & $\{x_4+x_1\}$\\ \hline
		$l_f<l_n$&$\{x_3+x_7\}$ &$\{x_1+x_4, x_5+x_2, x_6 +x_3\}$\\ \hline
	\end{tabular}
	
	\label{T3}
\end{table}


	\subsection{Rate Analysis}
\begin{figure}[h]
	\centering
	\includegraphics[width=6cm]{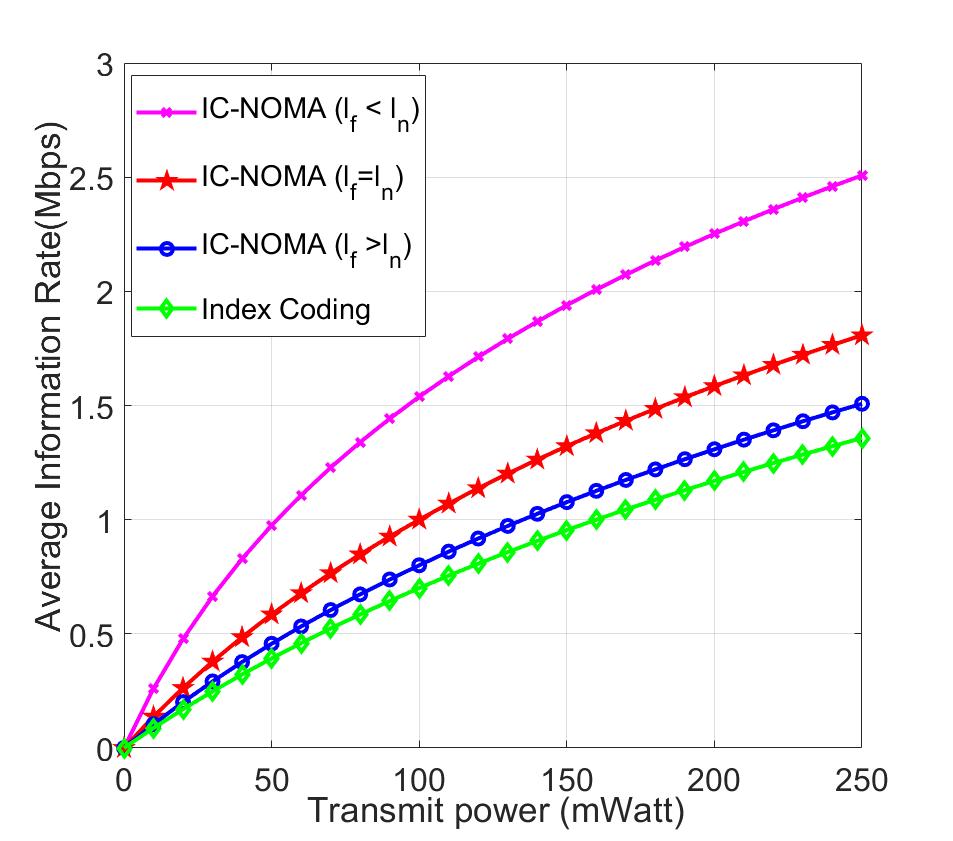}
	\caption{ Comparision of proposed system with conventional IC to show the improvement in average information rate for three different cases as given in Table \ref{T2}.}
	\label{figgg1}
\end{figure}

	Fig. \ref{figgg1} compares the average achievable rate of the proposed scheme as discussed in Section \ref{rate} considering a typical index coding scenario. In each IC-NOMA scenario, a total of $l^{\scriptscriptstyle IC-NOMA}$ transmission will be there with $l^{\scriptscriptstyle NOMA}$ NOMA transmissions and $l^{\scriptscriptstyle IC-NOMA} -l^{\scriptscriptstyle NOMA}$ IC transmissions. The average information rate, $R_{avg}$ for each case of IC-NOMA system is calculated as

	\begin{align}
&R_{avg} = \frac{l^{\scriptscriptstyle NOMA} R^{\scriptscriptstyle IN-NOMA} +(l^{\scriptscriptstyle IC-NOMA}-l^{\scriptscriptstyle NOMA})R^{\scriptscriptstyle IN-IC^{(i)} }}{l^{\scriptscriptstyle IC-NOMA}}.\nonumber
	\end{align}
We consider $R^{\scriptscriptstyle IN-IC^{(i)}}$ for $i=2$ and $i=3$ as the information rate of index coded transmission of IC-NOMA system for Case II and Case III respectively.
	The system performance varies with the lengths of index codes designed for far and near users. The plot shows that in IC-NOMA, the average information rate is high for $l_f < l_n$ when compared to the other other two cases. This is because, the information rate for the IC part of the IC-NOMA system for $l_f<l_n$ will be determined by the channel gains of near users.

	For the case $l_f>l_n$, both the far and near users need the NOMA part and IC part to satisfy their demands. The information rate for the IC part of the IC-NOMA system for $l_f > l_n$ will be determined by the channel gains of far users. Hence the information rate, in this case, will be less than that of the case $l_f < l_n$. 
	
	In IC-NOMA, for $l_f=l_n$, the far and near users satisfy their respective demands by NOMA transmissions only. Each NOMA transmission, in this case, would have a higher achievable rate than IC transmissions in the $l _f > l _n$ scenario. Hence IC-NOMA system with $l_f=l_n$ will outperform the scenario $l_f>l_n$ in terms of achievable information rate.

	\subsection{Transmission Power Analysis}
	In this, section we evaluate the average power requirement to achieve same information rate for IC and IC-NOMA scheme. The average power consumption $P_{avg}$ for each cases of IC-NOMA scheme is calculated as
		\begin{align}
	&P_{avg} =\frac{l^{\scriptscriptstyle NOMA} P_a +(l^{\scriptscriptstyle IC-NOMA}-l^{\scriptscriptstyle NOMA})P_b^{(i)}}{l^{\scriptscriptstyle IC-NOMA}} .\nonumber
	\end{align}
		\begin{figure}[]
		\centering
		\includegraphics[width=6cm, height=5.5cm]{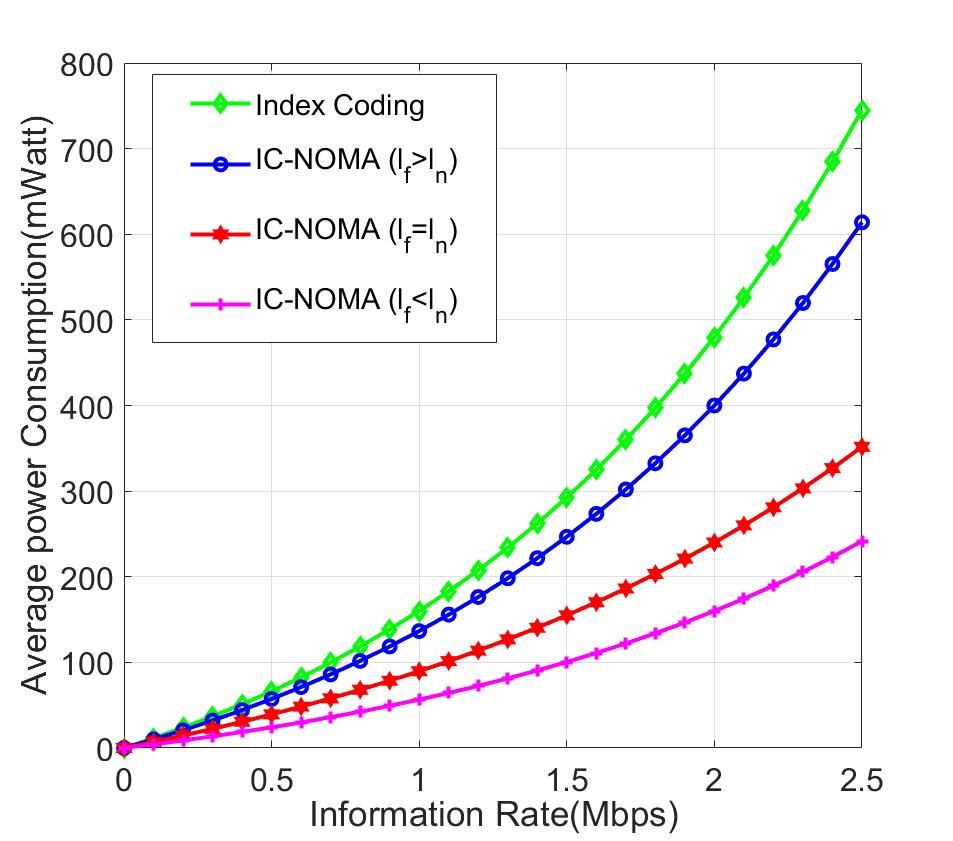}
		\caption{Comparision of the average power consumption for IC and three different cases of IC-NOMA system as given in Table \ref{T2}.}
		\label{figgg2}
	\end{figure}
The power per IC transmission is denoted as $P_b^{(i)}$ for $i=2$ and $i=3$ for Case II and Case III. Fig.\ref{figgg2} quantifies the analysis in Section \ref{ppower} for the example index coding scenario. It can be clearly seen that for achieving the same information rate, the proposed IC-NOMA system saves power when compared to conventional IC. In similar lines with the results of Fig. \ref{figgg1}, here also the power saving is the highest in the case where $l_f < l_n$. But here the performance of the $l_f =l_n$ case is more close to that of $l_f < l_n$ case since  this case requires only reduced number of NOMA transmissions.

	\subsection{QoS Requirement Analysis}
	This section graphically illustrate the power efficiency of proposed scheme compared to conventional IC to meet minimum the QoS requirements as discussed in Section \ref{qqos} by considering example scenario.
	
	From Section \ref{qqos}, it is clear that the total power needed for achieving the QoS requirements depends on the power allocation factor $\alpha$ and R.
	
\begin{figure}[H]
		\centering
		\includegraphics[width=6cm, height=5.5cm]{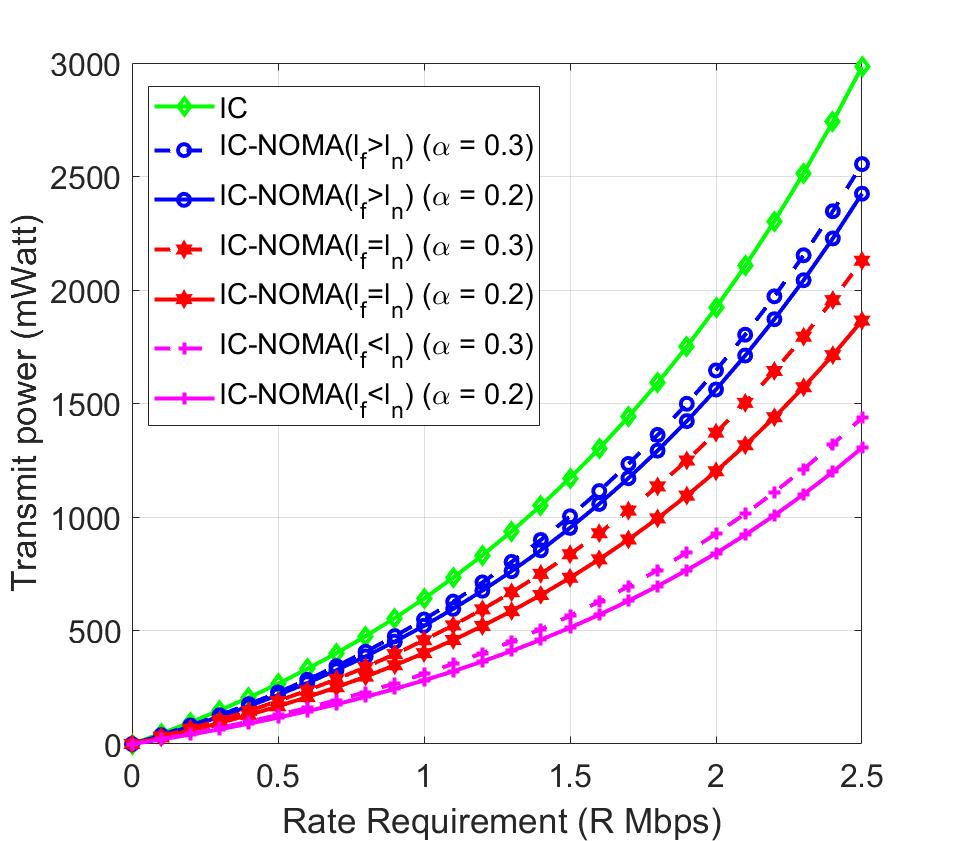}
		\caption{ The comparision of total power consumption for IC and three cases of IC-NOMA system as given in Table \ref{T2} to meet the QoS requirement (R) at individual user.}
		\label{fig:qosm2greater}
	\end{figure}

	The power requirement of conventional IC is independent of the power allocation factor, whereas the power requirement for NOMA transmissions in IC-NOMA varies with respect to $\alpha$. It is clear from the Fig. \ref{fig:qosm2greater} that just like the conventional NOMA, the optimal selection of power allocation factor is important for our proposed system to increase the power efficiency. 

	Eventhough  the value of optimal $\alpha$ depends on the channel gains of the users, the value of $\alpha$ generally varies between 0.2 and 0.3 for a conventional NOMA with 2 power levels as discussed in \cite{CN3}. Hence in this analysis we have quantified the power efficiency of proposed IC-NOMA for $\alpha$ values of 0.2 and 0.3. Fig.\ref{fig:qosm2greater} clearly shows that the proposed IC-NOMA provides  improved power efficiency while meeting the QoS requirements of various users.

	\section{Conclusion}
	In this paper, we proposed a spectral efficient transmission strategy called IC-NOMA for VANETs. IC-NOMA improves spectral efficiency by linearly superposing index coded packets in the power domain at the transmitter and decoding the desired data by exploiting the side information and SIC at the receiver. We showed that IC-NOMA requires a unique index code design to fit in with the NOMA scenario and developed an algorithm to design the same. The scenarios where IC-NOMA offer improved bandwidth efficiency compared to IC are studied. Based on this, the power saving of the proposed system is discussed by analytical studies.


\end{document}